\documentclass[preprint]{aastex}
\usepackage{mkfig}

\begin{document}

\title{\bf Detection of H$_{2}$ Emission from Mira~B in UV Spectra
  from the {\em Hubble Space Telescope}\altaffilmark{1}}

\author{Brian E. Wood,\altaffilmark{2} Margarita Karovska,\altaffilmark{3}
  and Warren Hack\altaffilmark{4}}

\altaffiltext{1}{Based on observations with the NASA/ESA Hubble Space
  Telescope, obtained at the Space Telescope Science Institute, which is
  operated by the Association of Universities for Research in Astronomy,
  Inc., under NASA contract NAS5-26555.}
\altaffiltext{2}{JILA, University of Colorado and NIST, Boulder, CO
  80309-0440.}
\altaffiltext{3}{Smithsonian Astrophysical Observatory, 60 Garden St., 
  Cambridge, MA 02138.}
\altaffiltext{4}{Space Telescope Science Institute, Baltimore, MD 21218.}

\begin{abstract}

     We present ultraviolet spectra of Mira's companion star from the Space
Telescope Imaging Spectrograph (STIS) instrument on board the {\em Hubble
Space Telescope} (HST).  The companion is generally assumed to be a white
dwarf surrounded by an accretion disk fed by Mira's wind, which dominates the
UV emission from the system.  The STIS UV spectrum is dominated by numerous,
narrow H$_{2}$ lines fluoresced by H~I Ly$\alpha$, which were not detected
in any of the numerous observations of Mira~B by the {\em International
Ultraviolet Explorer} (IUE).  The high temperature lines detected by IUE
(e.g., C~IV $\lambda$1550) still exist in the STIS spectrum but with
dramatically lower fluxes.  The continuum fluxes in the STIS spectra are
also much lower, being more than an order of magnitude lower than ever
observed by IUE, and also an order of magnitude lower than fluxes observed
in more recent HST Faint Object Camera objective prism spectra from 1995.
Thus, the accretion rate onto Mira~B was apparently much lower when STIS
observed the star, and this change altered the character of Mira~B's UV
spectrum.

\end{abstract}


\section{INTRODUCTION}

     Mira (o~Cet, HD~14386) is one of the most extensively observed variable
stars, representing the prototype of the ``Mira variable'' class of
pulsating stars.  These stars are red giants on the asymptotic giant branch
(AGB), with high mass loss rates and extended circumstellar envelopes.
Mira's pulsation properties are typical for long period Mira variables, with
large luminosity changes of about 7 magnitudes during its 332 day pulsation
cycle.  Its distance is 128~pc according to {\em Hipparcos} \citep{macp97}.

     Mira has a hot companion star, Mira~B (VZ~Cet).  The emission from
Mira~B is probably powered by a surrounding accretion disk fed by Mira's
massive wind.  The accretor is generally assumed to be a
white dwarf, although \citet{mj84} argue that the dearth of X-rays
from the Mira system suggests that the companion is more likely a low mass
main sequence star. The primary evidence for the accretion disk
is very broad UV emission lines observed by the {\em International
Ultraviolet Explorer} (IUE) (e.g., C~IV $\lambda$1550, Si~III] $\lambda$1892,
C~III] $\lambda$1909), which are believed to be from this rapidly rotating
disk \citep{ac85,dr85}.

     \citet{mk97} resolved the components of this binary at UV and
optical wavelengths using HST Faint Object Camera (FOC) images, and also
obtained the first spatially resolved spectra of each star individually.
They measured a separation between the components in 1995 of only
$0.578^{\prime\prime}$.  The FOC images and the ultraviolet objective
prism spectra provided for the first time detailed information on the
continuum and line emission from each component of the system, and allowed
a better measurement of the accretion luminosity of Mira B longward of
1500~\AA. The resolution of the objective prism spectra was, however,
insufficient for a detailed analysis of the line emission originating from
the accretion disk.

     In this paper, we present new high resolution
spectroscopic observations of Mira~B from the {\em Hubble Space Telescope}
(HST) to further study the nature of accretion onto this intriguing star.
In \S 2, we present the HST observations and compare them with previous UV
spectra from IUE and FOC, and in \S 3 we discuss our results.

\section{OBSERVATIONS AND RESULTS}

\subsection{Description of Observations}

     We observed Mira on 1999 August 2 using the Space Telescope Imaging
Spectrograph (STIS) instrument on board HST.  For a full description of
STIS, see \citet{rak98} and \citet{bew98}.  Unlike IUE,
the STIS instrument has apertures small enough to observe the two components
of the Mira~AB system separately.  Thus, there is no danger of any
contamination by Mira~A in our spectra.  The Mira~B
observations consisted of two moderate resolution UV spectra taken through the
$0.2^{\prime\prime} \times 0.2^{\prime\prime}$ aperture.  The first utilized
the E230M grating for a 600 s exposure of the 2303--3111~\AA\ wavelength
region.  The second used the E140M grating for an 1850 s exposure of the
1140--1735~\AA\ wavelength region.  The spectra were processed using the
standard HST/STIS pipeline software.  To test the accuracy of our wavelength
calibration, in the E140M spectrum we measured the centroid of the
geocoronal H~I Ly$\alpha$ emission peak to be at $+27.8\pm 0.4$ km~s$^{-1}$,
in perfect agreement with its expected location for that time of year
of $+27.9$ km~s$^{-1}$.

\subsection{The H$_{2}$ Lines}

     The top panel of Figure~1 shows the full E140M spectrum taken by
HST/STIS.  The spectrum is dominated by a very large number of narrow lines,
which were not seen in previous IUE observations of Mira~B.  We identified
them as being lines of molecular hydrogen. Furthermore, we believe
the lines are all fluoresced by the H~I Ly$\alpha$ line.  For example, the
0-4 R(0) and 0-4 P(2) lines identified in Figure~1 are pumped by the
0-2 R(0) line at 1217.2046~\AA, which lies within the very broad Ly$\alpha$
emission line from Mira~B.  Many other lines of the 0-x R(0) and 0-x P(2)
sequences are also detected with flux ratios roughly consistent with the
branching ratios \citep{ha93}.  This provides strong support for
the line identifications and for Ly$\alpha$ fluorescence being the source of
the emission.  Similarly, the 0-4 R(1) and 0-4 P(3) lines in Figure~1 are
pumped by the 0-2 R(1) 1217.6437~\AA\ line, and the 0-4 P(1) line is pumped
by 0-2 P(1) at 1219.3677~\AA.  

     Many Lyman band H$_{2}$ lines were first identified in the solar
spectrum.  The solar H$_{2}$ lines are fluoresced not only by Ly$\alpha$ but
also by transition region lines \citep{cj77,cj78}.  Lines of H$_{2}$
fluoresced by Ly$\alpha$ have by now been been detected in many other
types of astrophysical objects, including red giant stars
\citep{adm98,adm99}, T~Tauri stars \citep{ab81,jav00}, and Herbig-Haro
objects \citep{rds83,sc95}.

     The H$_{2}$ lines from Mira~B all have widths of about 20 km~s$^{-1}$,
well resolved by the $\sim 7$ km~s$^{-1}$ resolution of the E140M grating.
They are centered at a heliocentric velocity of $+57$ km~s$^{-1}$.  Various
observations of Mira's large, slowly expanding circumstellar
envelope suggest a systemic radial velocity of about $+56$ km~s$^{-1}$
\citep{pfb88,pp90,ej00}, consistent with our H$_{2}$ velocity.
A full analysis and list of the numerous H$_{2}$ Lyman band lines in
our FUV Mira~B spectrum will be presented in a future paper.

\subsection{Comparison with Previous Observations}

     The HST/STIS spectrum looks radically different from what we expected
to see based on previous IUE observations \citep{ac85,dr85}.
Previous IUE observations of Mira~B have instead shown a spectrum dominated
by a few lines commonly observed in far-UV (FUV) spectra of stars and other
astrophysical objects:  C~IV $\lambda$1550, O~I $\lambda$1300,
C~II $\lambda$1336, etc.  Figure~2 shows an example of one of the low
resolution IUE spectra (SWP7029) of this spectral region, with line
identifications based on previous analyses \citep{ac85,dr85}.

     In Figure~2, we rebinned and smoothed our STIS/E140M spectrum to
match the resolution of the IUE spectrum.  The deresolved STIS spectrum
shows peaks at 1335~\AA\ and 1400~\AA, which Figure~1 shows to be
predominantly from narrow H$_{2}$ emission lines.  In the IUE data, peaks at
these locations were previously identified with C~II and Si~IV lines,
respectively (with perhaps some O~IV] contribution to the 1400~\AA\ peak).
Were these features misidentified in the IUE spectra, and are they in fact
mostly H$_{2}$ emission in the IUE data as well as the STIS data?
In order to address this question we computed another deresolved STIS
spectrum for comparison with the IUE data, but only after removing all the
narrow H$_{2}$ lines.  The result is shown as a solid line in Figure~2.  The
peaks at 1335~\AA\ and 1400~\AA\ decrease dramatically but do not disappear
entirely, suggesting that there is some C~II and Si~IV emission in the STIS
data.  (The broad C~II emission near 1337~\AA\ can in fact be discerned in
both panels of Fig.~1.)  But what dominates these peaks in the IUE data,
C~II/Si~IV or H$_{2}$?

     The most prominent H$_{2}$ peak in the original STIS spectrum in Figure~2
(dotted line) that is not in danger of confusion with any other lines is the
peak at 1275~\AA, which completely disappears when the H$_{2}$ lines are
removed before deresolving the spectrum.  There are 32 short-wavelength,
low-resolution (SW-LO) spectra in the IUE archives.  In none of these spectra
is there a prominent peak at 1275~\AA.  Thus, we believe H$_{2}$ emission is
{\em not} an important contributor to any of the emission features observed
by IUE, including the peaks at 1335~\AA\ and 1400~\AA.  The STIS
spectrum of Mira~B is truly very different from anytime IUE observed it, and
this is not just a consequence of the superior resolution and S/N properties
of STIS compared with IUE.

     Further evidence of this is provided by the continuum fluxes of
our STIS spectra, which are well below those detected by IUE and
previous HST observations.  The fluxes of
the IUE spectrum in Figure~2 were divided by 20 to match the continuum
observed by STIS.  The IUE data set clearly shows some variability \citep[see][]{dr85}, but only variations of about a factor of 2 or
so.  The STIS continuum fluxes are an order of magnitude lower than any of
the IUE spectra.  Figure~3 shows that this result extends to the near-UV
(NUV) spectral region as well.  The top panel of Figure~3 compares a
rebinned, smoothed version of our STIS/E230M spectrum with a typical
long-wavelength, low-resolution (LW-LO) IUE spectrum (LWP18360).  Once
again, the continuum fluxes are different by a factor of 20, and once again
inspection of other LW-LO spectra in the IUE archive (31 total) shows this
to be a general result.  Comparison with the fluxes measured several years
earlier with the HST also shows a significant drop in emission.  The bottom
panel of Figure~3 shows that the STIS continuum fluxes are a factor of 10
lower than those observed in 1995 by an HST Faint Object Camera (FOC) PRISM
spectrum of Mira~B presented by \citet{mk97}.

     Emission line fluxes in our STIS spectra are also reduced relative to
IUE, but to varying degrees.  Most notable is the C~IV $\lambda$1550 line,
which dominates all the SW-LO IUE spectra, but is significantly weaker
relative to the continuum in our STIS data.  The C~IV flux observed by STIS
is roughly 100 times lower than observed by IUE (see Fig.~2).  This behavior
is different from that of cataclysmic variables, where C~IV emission tends
to become more prominent when the continuum decreases \citep{cwm87}.

     The optical emission from Mira B seems to also be low at the time of
our STIS observations.  In fact, we were not able to detect the companion 
in an image of Mira~AB taken by HST using the narrow band [O~III] filter of
STIS centered at 5007~\AA.  We estimate a lower limit for the apparent
magnitude of $V>11$.  Mira~B's optical variations in the range $V=9-12$ have
been suggested to have a possible period of 14 years \citep{ahj54,yy77}.

     For a final comparison with IUE data, in Figure~4 we compare the
Mg~II k line profile observed by STIS (with a vacuum rest wavelength of
2796.352~\AA) with a typical NUV, high resolution (LW-HI) observation
from IUE (LWP29795).  The two spectra are shown on a velocity
scale centered on the rest frame of the star.  The IUE fluxes are once again
divided by 20 to roughly match the STIS fluxes.  In the IUE profile, an
opaque absorption feature is observed between 0 and $-400$ km~s$^{-1}$ that
is indicative of a high speed wind from Mira~B.  Similar absorption has
been observed in optical H$\beta$ and H$\gamma$ lines \citep{yy77}.
Inspection of all 28 LW-HI IUE spectra available in the IUE archives
reveals that all of the Mg~II profiles have this absorption.  The extent of
the absorption does vary, sometimes extending to higher or lower velocities.
However, the wind absorption is always opaque out to at least
$-200$ km~s$^{-1}$ in the IUE data.  The STIS profile is different in that
the absorption is only opaque to $-50$ km~s$^{-1}$, and the absorption
between 0 and $-50$ km~s$^{-1}$ could conceivably be entirely interstellar.
There {\em is} clearly wind absorption on the blue side of the STIS Mg~II
profile, but the wind opacity is much less than in any of the IUE spectra.
This suggests that Mira~B's wind was much weaker at the time of the STIS
observations than when IUE observed the star.

\section{DISCUSSION}

     A fundamental question regarding the H$_{2}$ lines that we have
detected in the UV spectrum of Mira~AB is, where are they coming from?
The fluoresced H$_{2}$ molecules may be located in the cool
outer regions of the accretion disk surrounding Mira~B, or perhaps the
H$_{2}$ is in Mira~A's slow, cool wind, which is fluoresced as it approaches
the hot companion star and interacts with its faster wind (see below).
\citet{pfb88} estimated that hydrogen in Mira's wind should be
roughly 70\% molecular and 30\% atomic.  If the H$_{2}$ is in Mira's wind,
one might expect to see a velocity shift of the H$_{2}$ lines relative to
the systemic velocity, which we do not observe.  However, a shift is only
expected if Mira~B does not lie in the same plane of the sky as Mira~A.
Since Mira~B's orbit is not well known it is unclear what the geometry is.
Furthermore, the velocity of Mira~A's wind is very slow, with a terminal
expansion speed generally estimated to be in the $2-7$ km~s$^{-1}$ range
\citep[e.g.,][]{nr01}.  Thus, not much of a shift is to be expected and it
remains very possible that the H$_{2}$ lies within Mira's wind.

     The fluxes we observe in our HST/STIS observations of Mira~B's UV
spectrum are dramatically lower than ever observed by IUE.  Something
may have happened to reduce the accretion rate onto the companion, thereby
significantly lowering the accretion luminosity.  The lower accretion rate
leads to a lower mass loss rate from Mira~B, based on the Mg~II profiles
(see Fig.~4).

     It is not clear what could have caused the accretion rate to change.
Instabilities in the accretion disk could be responsible.  After all, more
luminous accretion systems, such as dwarf novae and cataclysmic variables,
show dramatic variability.  It also might have been caused by
inhomogeneities in the wind of Mira~A that feeds the accretion disk, as
there is evidence for substantial inhomogeneity in Mira~A's massive wind
\citep[e.g.,][]{bl97}.

     A final fundamental question is why the lower accretion rate is also
accompanied by more prominent emission from H$_{2}$ lines.  The lower
accretion luminosity and weaker wind might allow more H$_{2}$ molecules to
exist close to Mira~B, resulting in more fluorescence from Mira~B's broad
Ly$\alpha$ emission.  Future more detailed analysis and modeling of the
very rich H$_{2}$ fluorescence spectrum will shed some light on this issue.

\acknowledgments

We would like to thank J.\ Raymond for useful discussions and comments.
Support for this work was provided by NASA through grant number
GO-08298.01-99A from the Space Telescope Science Institute, which is operated
by AURA, Inc., under NASA contract NAS5-26555.  M.\ K.\ is a member of the
Chandra Science Center, which is operated under contract NAS8-39073,
and is partially supported by NASA.

\clearpage

\begin{figure}
\plotfiddle{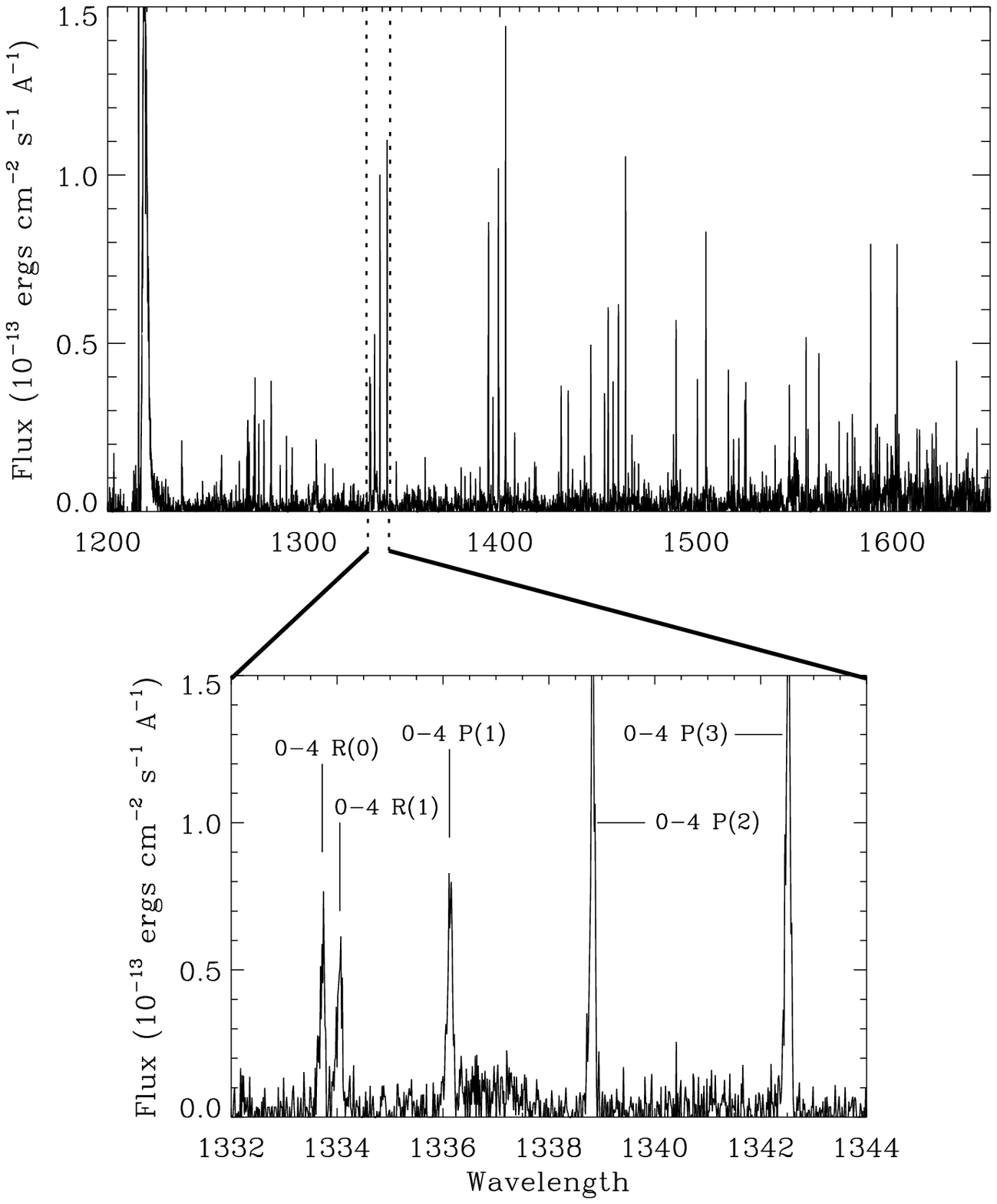}{4.5in}{0}{80}{80}{-270}{0}
\caption{A smoothed HST/STIS E140M spectrum of Mira~B, showing numerous
  narrow H$_{2}$ lines (top panel), and a blowup of a smaller
  region of this spectrum, showing five of these lines (bottom panel).}
\end{figure}

\clearpage

\begin{figure}
\plotfiddle{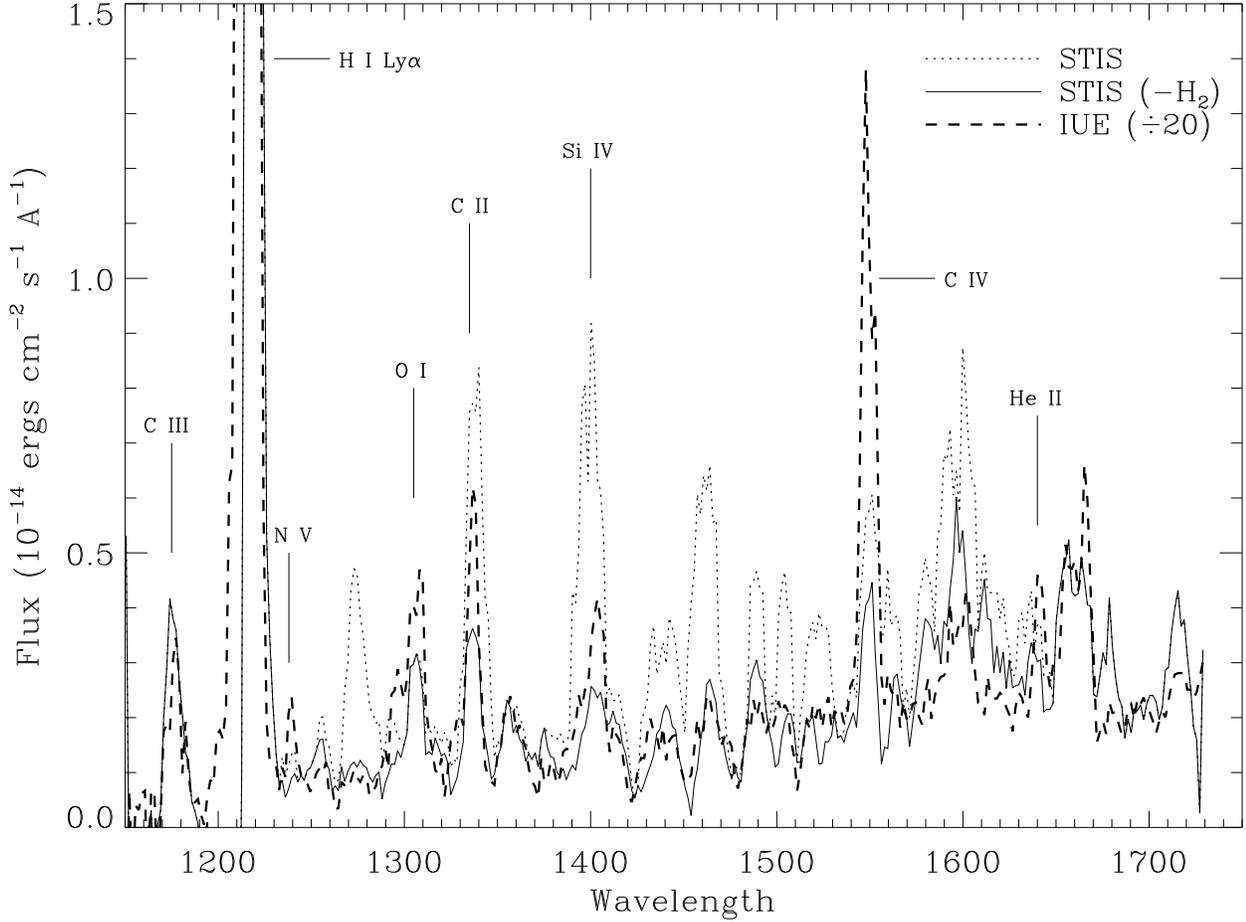}{3.5in}{90}{75}{75}{280}{0}
\caption{An IUE SWP-LO spectrum of Mira~B (dashed line), displayed with
  line identifications, is compared with our HST/STIS spectrum (dotted line),
  which has been rebinned and deresolved to match the resolution of the IUE
  spectrum.  We also show a version of the STIS spectrum deresolved after
  the removal of all the narrow H$_{2}$ lines (solid line).  Note that the
  IUE fluxes had to be reduced by a factor of 20 to match the continuum
  fluxes observed by HST/STIS.}
\end{figure}

\clearpage

\begin{figure}
\plotfiddle{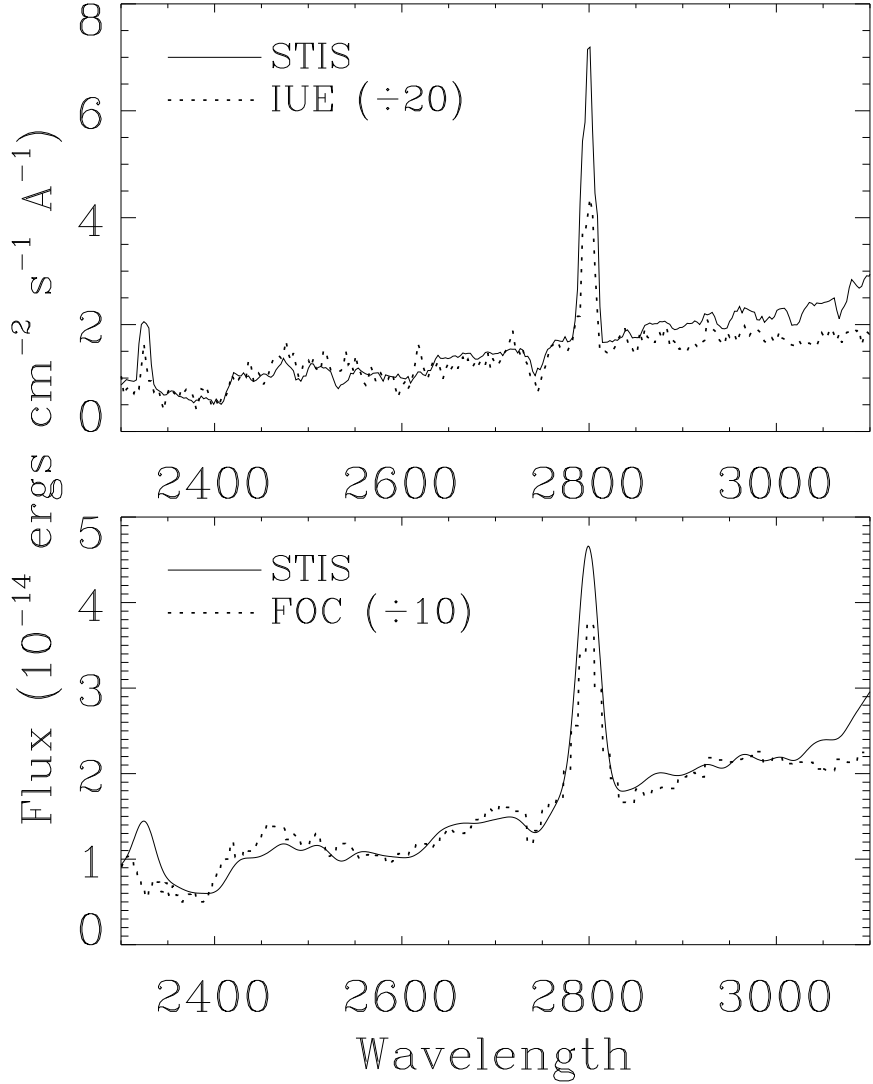}{4.5in}{0}{80}{80}{-270}{0}
\caption{Comparison of the HST/STIS E230M spectrum of Mira~B with
  previous observations from IUE (top panel) and HST/FOC (bottom panel).
  In both panels, the STIS spectrum is rebinned and deresolved to match the
  resolution of the other observation.  The peaks at 2325~\AA\ and 2800~\AA\
  are C~II] and Mg~II lines, respectively.  Note that the IUE and FOC fluxes
  had to be reduced by factors of 20 and 10, respectively, to match the STIS
  data.}
\end{figure}

\clearpage

\begin{figure}
\plotfiddle{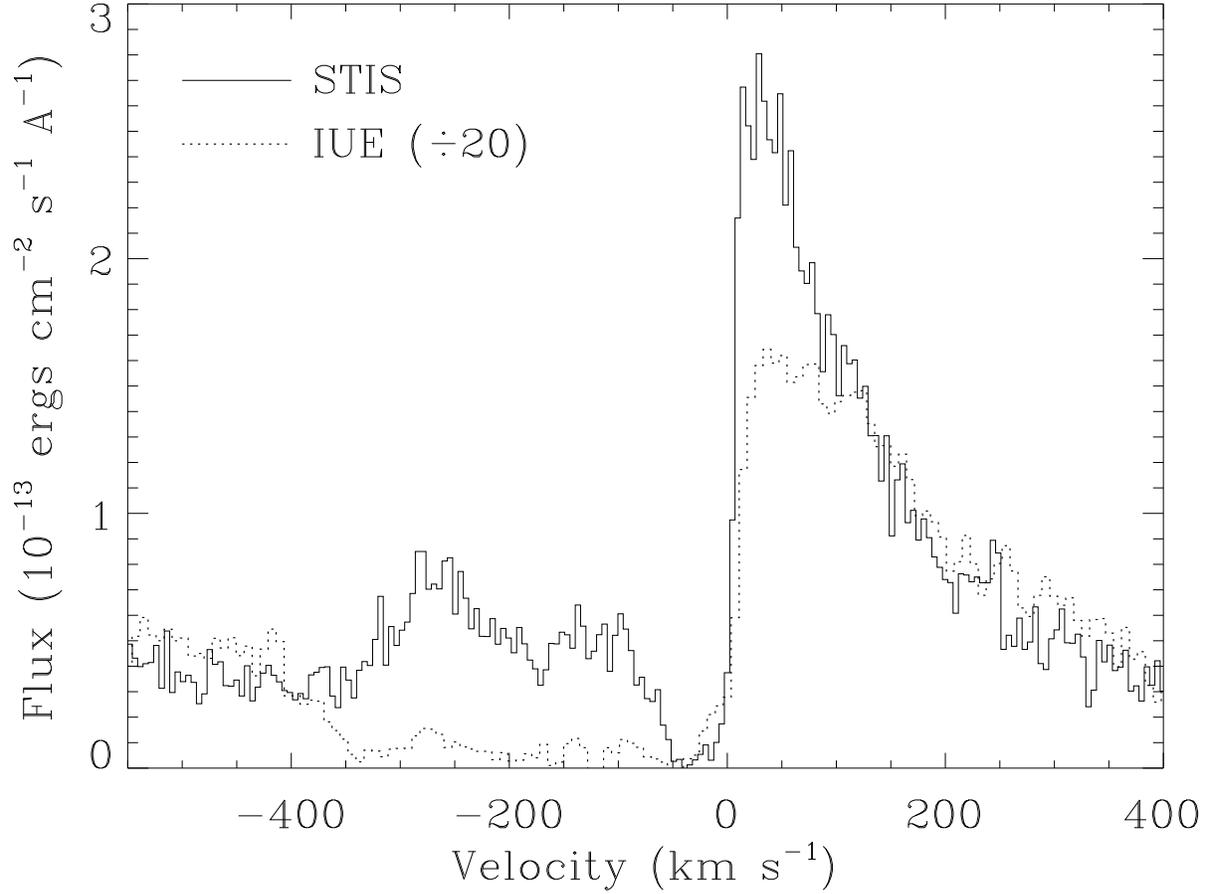}{3.5in}{90}{75}{75}{280}{0}
\caption{Comparison of the Mg~II k line profile observed by HST/STIS
  (solid line) and one observed by IUE (dotted line), shown on a velocity
  scale centered on the rest frame of the star.  The IUE fluxes are reduced
  by 20 to roughly match the STIS fluxes.  Note the larger wind opacity
  between 0 and $-400$ km~s$^{-1}$ in the IUE spectrum.}
\end{figure}

\end{document}